\pdfoutput=1
\documentclass[prd,preprint,superscriptaddress,preprintnumbers,nofootinbib]{revtex4}

\usepackage[dvips]{graphicx}
\usepackage{amsmath,amssymb,latexsym}
\usepackage{bm}
\usepackage{epsfig}

\newcommand{\postscript}[2]{\setlength{\epsfxsize}{#2\hsize}
   \centerline{\epsfbox{#1}}}

\newcommand{\mbh}{M_{\text{BH}}}

\usepackage[usenames,dvipsnames]{xcolor}
\definecolor{orange}{cmyk}{0,0.5,1,0}
\definecolor{rossoCP3}{cmyk}{0,.88,.77,.40}
\definecolor{graa}{rgb}{0.8,0.8,0.8}
\definecolor{blaa}{rgb}{0.2,0.2,0.6}

\begin{document}

\preprint{MPP-2022-60}
\preprint{LMU-ASC 24/22}


\title{\color{rossoCP3} The Dark Dimension, the Swampland, and the
  Dark Matter Fraction Composed of Primordial Black Holes}

\author{\bf Luis A. Anchordoqui}

\affiliation{Department of Physics and Astronomy,\\  Lehman College, City University of
  New York, NY 10468, USA
}

\affiliation{Department of Physics,\\
 Graduate Center, City University
  of New York,  NY 10016, USA
}

\affiliation{Department of Astrophysics,\\
 American Museum of Natural History, NY
 10024, USA
}

\author{\bf Ignatios Antoniadis}
\affiliation{Laboratoire de Physique Th\'eorique et Hautes \'Energies - LPTHE,\\
Sorbonne Universit\'e, CNRS, 4 Place Jussieu, 75005 Paris, France
}

\affiliation{SISSA, Via Bonomea 265, 34136, Trieste, Italy}
\affiliation{ICTP, Strada Costiere 11, 34151 Trieste, Italy}

\author{\bf Dieter\nolinebreak~L\"ust}

\affiliation{Max--Planck--Institut f\"ur Physik,\\  
 Werner--Heisenberg--Institut,
80805 M\"unchen, Germany
}

\affiliation{Arnold Sommerfeld Center for Theoretical Physics, \\
Ludwig-Maximilians-Universit\"at M\"unchen,
80333 M\"unchen, Germany
}

\begin{abstract}
 \vskip 2mm \noindent Very recently, it was suggested that 
 combining  the Swampland program with the smallness of
  the dark energy 
  and  confronting these ideas to experiment lead
  to the prediction of the
  existence of a single extra-dimension (dubbed the
dark dimension) with characteristic length-scale in the micron
range. We show that the rate of Hawking
radiation slows down for black holes perceiving the dark dimension and
discuss the impact of our findings in assessing the dark matter fraction
that could be composed of primordial black holes. We demonstrate that
for a species scale of ${\cal O}(10^{10}~{\rm GeV})$, an
all-dark-matter interpretation in terms of primordial black holes (PBHs)
should be feasible for 
masses in the range 
$10^{14} \alt \mbh/{\rm g} \alt 10^{21}$.  This
range is extended compared to that in the 4D theory by 3 orders of
magnitude in the low mass region. We 
also show that PBHs with $M_{\rm BH} \sim 10^{12}~{\rm g}$  could
potentially explain the well-known Galactic 511~keV gamma-ray line if
they make up a tiny fraction of the total dark matter density.  
\end{abstract}

\maketitle  

\section{Introduction}

The swampland program~\cite{Vafa:2005ui} seeks to demarcate the set of
four-dimensional effective field theories (EFTs) that can be coupled
to quantum gravity in a consistent way, e.g., the landscape of
superstring theory vacua, and discriminate
these theories from those that cannot, to strengthen the predictive
power of quantum gravity in general, and superstring theory in particular. This
is accomplished by enumerating criteria that an EFT must fulfill so as
to be in the landscape, rather than be relegated to the ``swampland.''
These criteria have evolved to some set of conjectures, which can be
used as new guiding principles to construct compelling UV-completions
of the Standard Model (SM). Moreover, the UV constraints on IR physics
have led to a shift in the way we approach cosmology model building. There are many swampland conjectures in the market, indeed too many to be listed here and readers are referred to comprehensive reviews~\cite{Palti:2019pca,vanBeest:2021lhn}.

It was argued some time ago that {\it the cosmological
  hierarchy problem}, i.e. the
smallness of dark energy in Planck units
($\Lambda \sim 10^{-122} M_{\rm Pl}^4$),
can be explained statistically~\cite{Bousso:2000xa} or even anthropically~\cite{Weinberg:1987dv,Susskind:2003kw,Schellekens:2006xz}
by the huge number of vacua in the string landscape.
However very recently, it was suggested~\cite{Montero:2022prj} that by
a combination of {\it the cosmological
  hierarchy problem} and {\it the distance
conjecture}
one
naturally ends up in a peculiar corner of the string landscape, namely  with
 a single extra-dimension characterized by a length-scale in the micron
range.\footnote{Large extra dimension scenarios were originally introduced to solve the electroweak hierarchy problem 
\cite{Antoniadis:1990ew,Arkani-Hamed:1998jmv,Antoniadis:1998ig}.}

The   distance
conjecture  predicts the appearance of infinite towers of states that become
light and imply a breakdown of the EFT in the infinite distance limits
in moduli space~\cite{Ooguri:2006in}. Stated in the form of the
anti-de Sitter (AdS)
distance conjecture~\cite{Lust:2019zwm}, it  suggests that
there should be an infinite tower of states, whose mass is related to
the magnitude of the cosmological constant. More precisely, the mass scale $m$
behaves as $m \sim |\Lambda|^\alpha$, as the negative AdS vacuum energy
 $\Lambda \to 0$, with $\alpha$ a
 positive constant of ${\cal O} (1)$.  In~\cite{Lust:2019zwm} also some implications of the AdS distance conjecture 
 for de Sitter space were discussed, namely, when assuming
 this scaling behavior  to  hold in dS (or
 quasi dS) space with a positive cosmological constant,  approaching $\Lambda = 0$ will also lead to an unbounded number of massless modes.

 To each tower we can associate two mass scales: $m$,
 which is the mass scale of states in the tower, and $\hat M$, which is the
 scale local EFT description breaks down. The latter is the so-called
 ``species scale''~\cite{Dvali:2007hz,Dvali:2007wp}  that corresponds to
 the Planck scale of the higher dimensional theory, 
   $\hat M =  m^{n/(n+2)} M_{\rm Pl}^{2/(n+2)}$, where $n$ is the number
   of effective dimensions decompactifying. Requiring the experimental
 bound on deviations from Newton's gravitational inverse-square law~\cite{Lee:2020zjt} to
 be consistent with the theoretical bound from the swampland
 conjectures leads to $\alpha = 1/4$ and so the mass scale of the KK
 modes in the tower is estimated to be $m \sim \lambda^{-1}
 \Lambda^{1/4}$. Consistency with neutron star heating~\cite{Hannestad:2003yd} yields $n=1$~\cite{Montero:2022prj},
 whereas consistency with the sharp cutoff observed in the cosmic ray
 spectrum requires $\lambda \sim
 10^{-3}$~\cite{Anchordoqui:2022ejw}. On the whole, swampland
 considerations combined with observational data lead to the prediction of a single extra
mesoscopic dimension of length $R \sim \lambda \Lambda^{-1/4} \sim
1~\mu {\rm m}$, where $\Lambda^{1/4} = 2.31~{\rm meV}$. This extra
dimension, nicknamed the dark dimension,
opens up at the scale $m$ of the tower, where physics must be
described by an EFT in higher dimensions up to the species
scale $\hat M \sim 10^{10}~{\rm GeV}$. 

In this paper we study some phenomenological aspects of black
holes perceiving the dark dimension and, in the spirit
of~\cite{Argyres:1998qn}, we investigate the impact of
these higher dimensional objects in assessing the fraction of dark
matter that could be composed of primordial black holes, $f_{\rm PBH}$. The layout of
the paper is as follows. We begin in Sec.~\ref{sec:2} with an overview
of existing limits on $f_{\rm PBH}$. This includes constraints from
the isotropic photon backgrounds, observations of the cosmic microwave
background (CMB), and measurements of the positron density at the Galactic bulge. In Sec.~\ref{sec:3} we first reexamine the rate of Hawking radiation of 5-dimensional black
holes with the new species scale in
mind, and after that we confront our predictions to experiment. In
Sec.~\ref{sec:4} we reexamine within the 5D theory whether black holes
evaporating right now could
be responsible for the excess of 511~keV photons observed from the
inner Galaxy by the SPI spectrometer on board the INTEGRAL
satellite~\cite{Knodlseder:2005yq,Jean:2005af}. In Sec.~\ref{sec:5} we
discuss the possibility for primordial black holes to grow
via the accretion. The
paper wraps up with some conclusions presented in Sec.~\ref{sec:6}. Before proceeding, we pause to note that a specific realization of the
model proposed in~\cite{Montero:2022prj} should guarantee that the SM
interacts with the extra dimension only gravitationally, while the
cosmological constant scale is fixed by the size of the extra
dimension. Although {\it a priori} this does not seem obvious at all,
we continue here on the assumption that such a realization can indeed
emerge from string theory.

\section{Primordial Black Holes as a Dark Matter Candidate}
\label{sec:2}

\begin{figure}[tpb]
\postscript{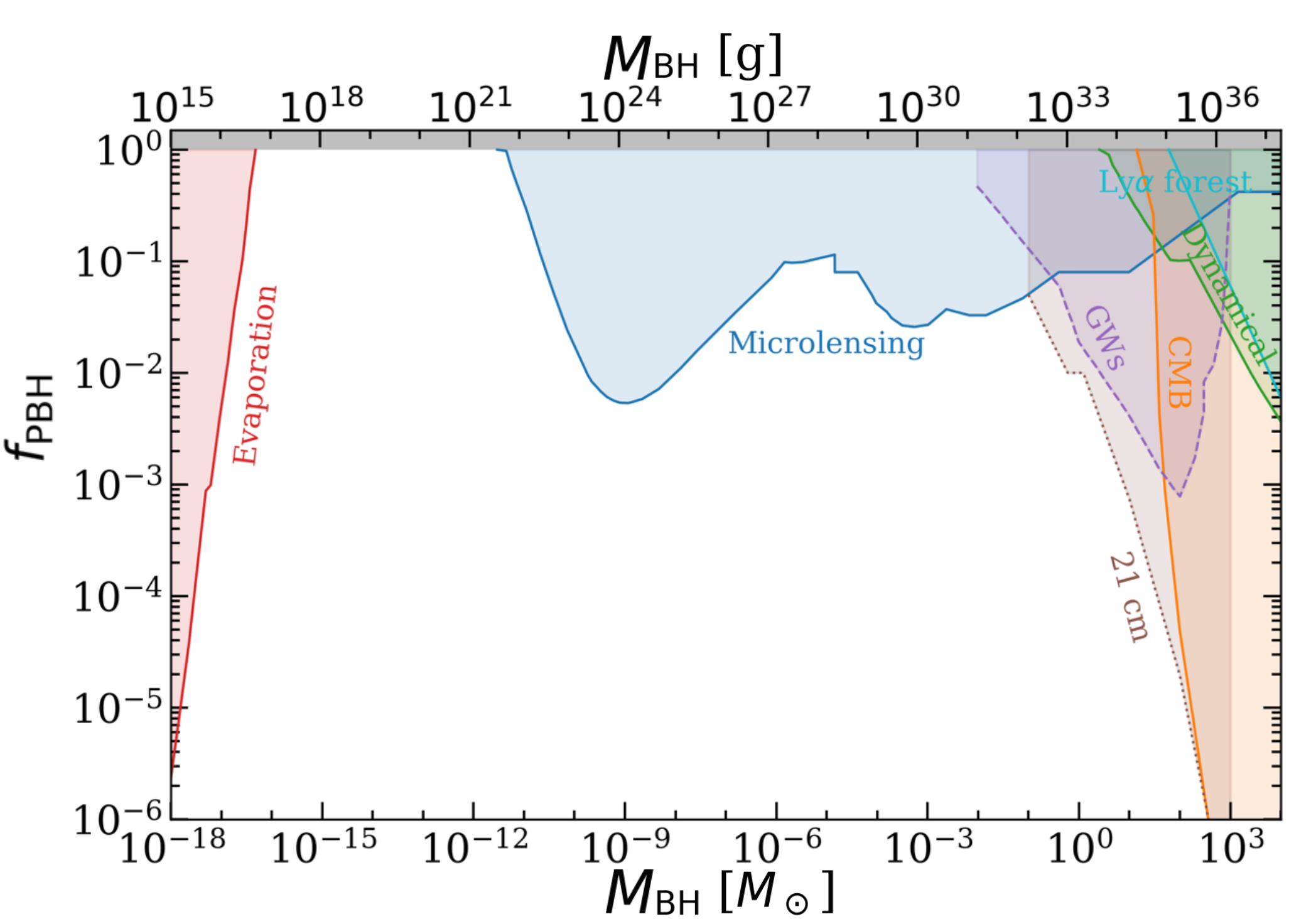}{0.8}
\caption{(Taken from~\cite{Villanueva-Domingo:2021spv}.) Compilation of contraints on $f_{\rm PBH}$ as a function of the PBH mass $M_{\rm BH}$, assuming a monochromatic mass
  function. The different probes considered are: the impact of PBH
  evaporation (red) on the extragalactic $\gamma$-ray background~\cite{Carr:2009jm} and on the CMB spectrum~\cite{Clark:2016nst}; non-observation of microlensing events (blue)
  from the MACHO~\cite{Macho:2000nvd}, EROS~\cite{EROS-2:2006ryy},
  Kepler~\cite{Griest:2013aaa}, Icarus~\cite{Oguri:2017ock}, OGLE~\cite{Niikura:2019kqi} and Subaru-HSC~\cite{Croon:2020ouk}
  collaborations; PBH accretion signatures on the CMB (orange),
  assuming spherical accretion of PBHs within halos~\cite{Serpico:2020ehh}; dynamical constraints, such as
  disruption of stellar systems by the presence of PBHs (green), on
  wide binaries~\cite{Monroy-Rodriguez:2014ula}  and on ultra-faint dwarf
  galaxies~\cite{Brandt:2016aco}; power spectrum from the
  Ly$\alpha$ forest (cyan)~\cite{Murgia:2019duy}; merger rates
  from gravitational waves (purple), either from individual mergers~\cite{Kavanagh:2018ggo,LIGOScientific:2019kan}
  or from searches of
  stochastic gravitational wave background~\cite{Chen:2019irf}.
 Gravitational wave limits, denoted
  by dashed lines, are model dependent~\cite{Boehm:2020jwd}. The dotted brown line corresponds to forecasts
  from the 21 cm power spectrum with SKA
  sensitivities~\cite{Mena:2019nhm} and from 21 cm forest
  prospects~\cite{Villanueva-Domingo:2021cgh}.}
\label{fig:1}
\end{figure}

It has long been suspected that black holes could emerge from the
collapse of large amplitude fluctuations in the very early
universe~\cite{Zeldovich:1967lct,Hawking:1971ei,Carr:1974nx,Carr:1975qj}. Although
the mass spectrum of these primordial black holes (PBHs) is not set in
stone, on cosmological scales they would behave like a typical cold
dark matter particle. Actually, the idea that PBHs could be
interesting dark matter candidates dates back at least as far as
1975~\cite{Chapline:1975ojl}, with punctuated revivals of activity
following the microlensing searches for massive compact halo objects
(MACHOs) in 1997~\cite{MACHO:1996qam}, and the LIGO/Virgo detections
of merging binary black holes in
2016~\cite{LIGOScientific:2016aoc}. The first microlensing searches
suggested that dark matter could be composed of MACHOs with mass
$\sim 0.5M_\odot$, which is the expected mass scale for PBHs produced
during the quark-hadron phase
transition~\cite{Jedamzik:1998hc}. However, more recent observations
exclude significant contributions of MACHOs to dark matter over most
of the plausible mass range~\cite{MACHO:2000qbb,Macho:2000nvd,EROS-2:2006ryy,Griest:2013aaa,Oguri:2017ock,Niikura:2019kqi,Croon:2020ouk}. The
question of whether the LIGO/Virgo merger events correspond to black
holes of astrophysical or primordial origin is still under
debate~\cite{Bird:2016dcv,Sasaki:2016jop,Hall:2020daa}, and a mixed
population may also be compatible with
observations~\cite{DeLuca:2021wjr}. However, data suggest that the
binary black hole merging rate is incompatible with an all-dark-matter
scenario and that PBHs could only contribute to less than 1\%
of the total dark matter~\cite{LIGOScientific:2019kan,Wong:2020yig}.

The mass distribution of PBHs is usually characterized by the mass function
\begin{equation}
\psi (M_{\rm BH}) = \frac{M_{\rm BH}}{\rho_{\rm CDM}} \ \frac{dn_{\rm
    PBH}}{dM_{\rm BH}} \,, 
\label{psiM}
\end{equation}
where $M_{\rm BH}$ is the black hole mass, $dn_{\rm PBH}$ is the number density of PBHs within the mass
range $(M_{\rm BH}, M_{\rm BH} + dM_{\rm BH})$, and $\rho_{\rm CDM}$
is the energy density of cold dark matter~\cite{DeLuca:2020ioi}. Integrating $\psi (M_{\rm
  BH})$ gives the total fraction of dark matter in PBHs,
\begin{equation}
f_{\rm PBH} \equiv \frac{\rho_{\rm PBH}}{\rho_{\rm CDM}} = \int \psi
(M_{\rm BH}) \, dM_{\rm BH} \,, 
\end{equation}
where $\rho_{\rm PBH} = \int M_{\rm BH} \, dn_{\rm PBH}$ is the energy
density of PBHs. If all of the dark matter were made of PBH,
we would have $f_{\rm PBH} = 1$. A compilation of upper
limits on the dark matter fraction that can be composed of PHBs is shown in
Fig.~\ref{fig:1}.

The question we want to address herein is whether primordial black
holes perceiving the dark dimension could ameliorate the constraints
on $f_{\rm PBH}$ shown in Fig.~\ref{fig:1}. In terms of the size of
the extra dimension $R$ and the string length $l_s$, we can
distinguish three definite regimes for the black hole horizon $r_s$:
{\it (i)}~$r_s > R$, where the theory looks like 4D and
$r_s \simeq M_{BH}$; {\it (ii)}~$l_s < r_s < R$, where the black hole
perceives the higher dimensional space; {\it (iii)}~$r_s < l_s$, where
the black hole turns into a string
state~\cite{Horowitz:1996nw,Horowitz:1997jc}. In our analysis, only
the regime {\it (ii)} will be relevant, i.e. we will study black holes
that are smaller than the size of the extra dimension, but larger than
the string size.

\section{Radiation Time-Scale of Five-Dimensional Black Holes}
\label{sec:3}

PBHs will Hawking evaporate, provided the semiclassical approximation
is valid. The average number~\cite{Hawking:1975vcx,Hawking:1974rv} and the
probability distribution of the
number~\cite{Parker:1975jm,Wald:1975kc,Hawking:1976ra} of outgoing particles in
each mode obey a thermal spectrum, with temperature~\cite{Anchordoqui:2001cg} 
\begin{equation}
T_{\rm BH} = \frac{n +1}{4\,\pi\,r_s} 
\label{TBH}
\end{equation}
and entropy
\begin{equation}
  S = \frac{4 \, \pi\,\mbh\,r_s}{n+2} \,,
\end{equation}
where
\begin{equation}
r_s(\mbh) =
\frac{1}{M_{{\rm Pl},n}}
\left[ \frac{\mbh}{M_{{\rm Pl},n} } \,\, \frac{2^n \pi^{(n-3)/2}\Gamma({n+3\over 2})}{n+2}
\right]^{1/(1+n)}\,,
\label{Sch}
\end{equation}
is the radius of a $(4+n)$-dimensional Schwarzschild black hole,
\begin{equation}
M_{{\rm Pl},n} = \left(\frac{m^n \ M_{\rm Pl}^2}{8 \pi}\right)^{1/(n+2)} \,, 
\end{equation}
and where 
$\Gamma(x)$ is the
Gamma function~\cite{Myers:1986un}.

The black hole, however, produces an effective potential barrier
surrounding the event horizon that backscatters part of the outgoing
radiation, making alterations to the Planckian spectrum. The black hole absorption
cross section, $\sigma_s$ (a.k.a. the greybody factor), depends on:
{\it (i)}~the spin $s$ of particle being emitted, {\it (ii)}~the
particle's energy $Q$, and {\it (iii)}~$\mbh$~\cite{Page:1976df}. At high frequencies ($ Q r_s \gg 1$)
the greybody factor for each kind of particle must approach the
geometrical optics limit. The integrated power emission is reasonably
well approximated taking such a high energy limit. In our calculations
we adopt the geometric optics
approximation, where the black hole acts as a perfect absorber of a
slightly larger radius, with emitting area given by
\begin{equation}
A_{4 \subset 4+n} = 4 \pi   \left(\frac{n+3}{2} \right)^{2/(n+1)} \frac{n+3}{n+1} \,  r_s^2 \,\,.
\end{equation}
Within this framework, we can conveniently write the greybody factor
as a dimensionless constant normalized to the black hole surface area
seen by the SM fields $\Gamma_s = \sigma_s/A_{4 \subset 4+n},$ such that
$\Gamma_{s=0} = 1$, $\Gamma_{s=1/2} \approx 2/3$, and $\Gamma_{s=1}
\approx 1/4$~\cite{Han:2002yy}.

All in all, a black hole emits
particles with initial total energy between $(Q, Q+dQ)$ at a rate
\begin{equation}
\frac{d\dot{N}_i}{dQ} = \frac{\sigma_s}{8 \,\pi^2}\,Q^2 \left[
\exp \left( \frac{Q}{T_{\rm BH}} \right) - (-1)^{2s} \right]^{-1}
\label{rate}
\end{equation}
per degree of particle freedom $i$. The change of variables $u=Q/T,$ 
brings
Eq.~({\ref{rate}) into a more familiar form,
\begin{equation}
\dot{N}_i = f \frac{ \Gamma_s}{32 \,\pi^3}\,
\frac{(n+3)^{(n+3)/(n+1)} (n+1)}{2^{2/(n+1)}} \, T_{\rm BH} \,
\int \frac{u^2}
{e^u - (-1)^{2s}} \,du.
\label{rate3}
\end{equation}
This expression can be easily integrated using 
\begin{equation}
\int_{0}^\infty \frac{z^{n-1}}{e^z - 1}\, dz = \Gamma(n)\, \zeta (n)
\end{equation}
and
\begin{equation}
\int_{0}^\infty \frac{z^{n-1}}{e^z + 1}\, dz = \frac{1}{2^n}\, (2^n -2)\,
\Gamma(n)\, \zeta (n)\,\,,
\end{equation}
yielding
\begin{equation}
\dot{N_i} = f \frac{\Gamma_s}{32\,\pi^3} \,
\frac{(n+3)^{(n+3)/(n+1)}\,(n+1)}{2^{2/(n+1)}} \,\Gamma(3) \, \zeta(3) \,T_{\rm BH}\,,
\end{equation}
where $\zeta(x)$ is the Riemann zeta function
and $f=1$ ($f=3/4$) for bosons (fermions). Therefore, the black hole emission
rate is found to be
\begin{equation}
\dot{N}_i \approx 3.7 \times 10^{21}\, \frac{(n+3)^{(n+3)/(n+1)}}{2^{2/(n+1)} \,(n+1)^{-1}}\,
\left(\frac{T_{\rm BH}}{{\rm GeV}}\right)\,\, {\rm s}^{-1} \,\,,
\end{equation}
\begin{equation}
\dot{N}_i \approx 1.8 \times 10^{21}\, \frac{(n+3)^{(n+3)/(n+1)}}{2^{2/(n+1)} \,(n+1)^{-1}}\,
\left(\frac{T_{\rm BH}}{{\rm GeV}}\right)\,\, {\rm s}^{-1} \,\,,
\end{equation}
\begin{equation}
\dot{N}_i \approx 9.2 \times 10^{20}\,\frac{(n+3)^{(n+3)/(n+1)}}{2^{2/(n+1)}
\,(n+1)^{-1}}\,
\left(\frac{T_{\rm BH}}{{\rm GeV}}\right)\,\, {\rm s}^{-1} \,\,,
\end{equation}
for particles with $s = 0,\, 1/2, \,1,$ respectively~\cite{Anchordoqui:2002cp}.

At any given time, the rate of decrease in the black hole mass is just
the total power radiated
\begin{equation}
\frac{\dot M_{\rm BH}}{dQ} = - \sum_i c_i \,\frac{\sigma_s}{8 \pi^2}\, 
\frac{Q^3}{e^{Q/T_{\rm BH}} - (-1)^{2s}} \,\,,
\end{equation}
where $c_i$ is the number of internal degrees of freedom of particle 
species $i.$ A straightforward calculation yields
\begin{equation}
  \dot M_{\rm BH} = - \sum_i c_i\,\, \tilde f\,\, \frac{ \Gamma_s}{32 \,\pi^3}\,
\frac{(n+3)^{(n+3)/(n+1)} (n+1)}{2^{2/(n+1)}} \, \Gamma(4) \, \zeta (4) \,\,T_{\rm BH}^2,
\label{dife}
\end{equation}
where $\tilde f = 1 \,\, (\tilde f = 7/8)$ for bosons
(fermions). Herein, we assume that the effective high energy theory
contains approximately the same number of modes as the SM
(i.e. $c_{s=0} = 1$, $c_{s=1/2} = 90,$ and $c_{s=1} = 27$) and we
neglect the effect of graviton emission.\footnote{A point worth noting
  at this juncture is that at first sight it may appear that the KK
  modes must dominate Hawking radiation because there are a large
  number -- ${\cal O} (R/r_s)^2$ -- light modes with masses below the
  $T_{\rm BH}$ scale. However, as noted in~\cite{Emparan:2000rs} it is
  incorrect to think of the individual KK modes of the bulk graviton
  as massive spin two fields on the brane with standard (minimal)
  gravitational couplings. Rather, since the KK modes are excitations
  in the full transverse space, their overlap with the small
  (higher-dimensional) black holes is suppressed by the geometric
  factor $(r_s/R)^2$ relative to the brane fields. Thus, the geometric
  suppression precisely compensates for the enormous number of modes,
  and the total contribution of all KK modes is only the same order as
  that from a single brane field.}

\subsection{Black hole radiation rate for $\bm{n=0}$}

The rate of Hawking radiation is estimated to be
\begin{equation}
\left. \frac{dM_{\rm BH}}{dt} \right|_{n=0} \simeq - 9 \times 10^{73}~{\rm GeV}^4 \frac{1}{M_{\rm BH}^2} \,\,.
\label{amateur}
\end{equation}
Ignoring accretion and thresholds, i.e., assuming that the mass of the black hole evolves
according to Eq.~(\ref{amateur}) during the entire process of evaporation, we
can obtain an estimate for the lifetime of the black hole,
\begin{equation}
\tau_{\rm BH}^{n=0} \simeq 1 \times 10^{-74}~{\rm GeV}^{-4}\,\,\int M_{\rm BH}^2\,\, dM_{\rm BH} \,\,.
\label{entree}
\end{equation}           
Using $\hslash = 6.58 \times 10^{-25}~{\rm GeV\, s},$ Eq.~(\ref{entree})
can then be rewritten as
\begin{eqnarray}
\tau_{\rm BH}^{n=0}  \simeq  1.6 \times 10^{-35} \,\, 
(M_{\rm BH}/{\rm g})^3~{\rm yr} \,\, .
\end{eqnarray}
Note that a black hole with $M_{\rm BH} \sim 5 \times 10^{14}~{\rm
  g}$ will have a lifetime of about 2 Gyr, comparable to the age of
the Universe~\cite{MacGibbon:2007yq,MacGibbon:2015mya}. Therefore, PBHs with
$\mbh \alt 5 \times 10^{14}~{\rm g}$  cannot form part of the observed
dark matter density. PBHs with masses small enough, but still {\it
  alive} in the Universe, should emit strong photon and cosmic
ray backgrounds which could be
observed~\cite{Laha:2020ivk,Iguaz:2021irx}. Null results from detection of these backgrounds
exclude an all dark matter interpretation in terms of PBHs for masses
$\mbh \alt 10^{17}~{\rm g}$. As shown in
Fig.~\ref{fig:1},  the allowed mass
range for the PHB dark matter
interpretation   is $10^{17} <\mbh/{\rm g} \alt 10^{21}$.

\subsection{Black hole radiation rate for $\bm{n=1}$}

\begin{figure}[tpb]
\postscript{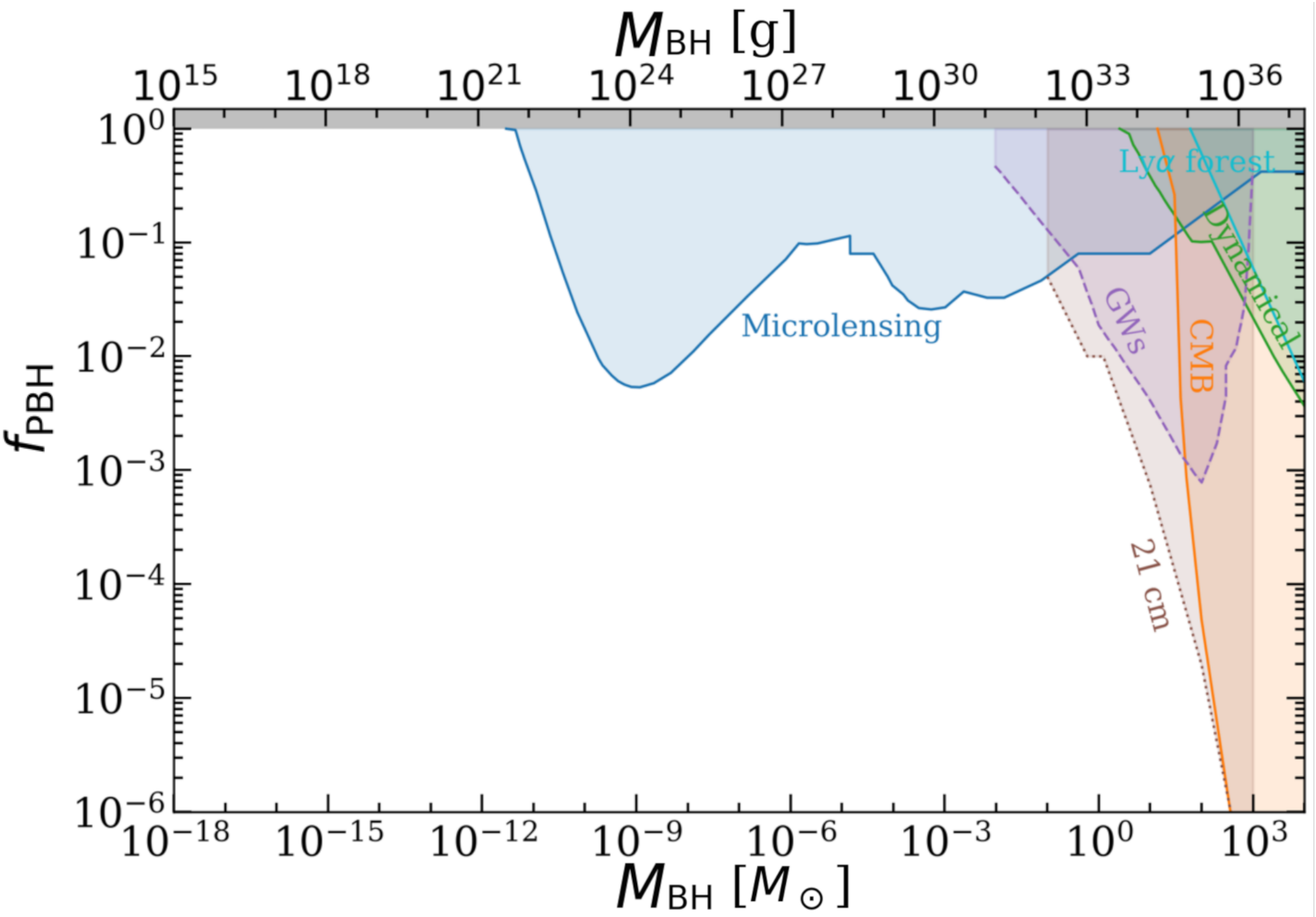}{0.8}
\caption{Compilation of constraints on $f_{\rm PBH}$ in the 5D theory as a function of
  the PBH mass $M_{\rm BH}$, assuming a monochromatic mass
  function. See Fig.~\ref{fig:1} for details of the different probes considered.}
\label{fig:2}
\end{figure}

We now assume the black hole can be treated as a flat
$(4+n)$-dimensional object. This assumption is valid for extra
dimensions that are larger than the Schwarzschild
radius~\cite{Argyres:1998qn}. For
$M_{{\rm Pl},n} \sim 10^{10}~{\rm GeV}$, the rate of Hawking radiation
is estimated to be
\begin{equation}
\left. \frac{dM_{\rm BH}}{dt} \right|_{n=1} \simeq - 4 \times 10^{29}~{\rm GeV}^3 \frac{1}{M_{\rm BH}} \,\,,
\label{amateur2}
\end{equation}
and so
\begin{equation}
\tau_{\rm BH}^{n=1} \simeq 3 \times 10^{-30}~{\rm GeV}^{-3}\,\,\int M_{\rm BH}\,\, dM_{\rm BH} \,\,,
\label{entree2}
\end{equation}           
which implies
\begin{eqnarray}
\tau_{\rm BH}^{n=1}  \simeq  9 \times 10^{-15} \,\, 
(M_{\rm BH}/{\rm g})^2~{\rm yr} \,\,.
\end{eqnarray}
For $n=1$, a black hole lives longer than a usual $n=0$ black hole of the same mass. A black hole with
$\mbh \sim 5 \times 10^{11}~{\rm g}$ has a lifetime approximately equal
to the age of the Universe.

As shown in Fig.~\ref{fig:1}, 
 PBHs in the 4D theory with $10^{15} \alt \mbh/{\rm g} \alt 10^{17}$ are
incompatible with an all-dark-matter interpretation. However, 
black holes sensing the dark dimension slow down the
Hawking radiation. For a given mass, the black hole
lives longer in the 5D theory than in the 4D theory. This implies that
black holes sensing the extra mesoscopic dimension emit less particles
and so the limits from isotropic photon
backgrounds~\cite{Laha:2020ivk,Iguaz:2021irx}, CMB observations~\cite{Poulin:2016anj,Poulter:2019ooo}, and measurements of the positron density at the Galactic bulge~\cite{Laha:2019ssq,DeRocco:2019fjq} can be relaxed. Via a
direct comparison of our calculations with the limits shown in 
Fig.~\ref{fig:1} we can 
conclude that an all-dark-matter interpretation in terms of PBHs in
the 5D theory should be feasible for 
$10^{14} \alt \mbh/{\rm g} \alt 10^{21}$, thus extending the allowed
mass range by 3 orders of magnitude.

Three observations are in order: {\it (i)}~The temperature of PBHs
evaporating today may not be enough to emit all the SM degrees of
freedom. Herein we are interested in the order of magnitude estimate and we
take this to fall within errors. {\it(ii)}~The lack of femtolensing
detection in the gamma-ray burst data have been interpreted as
evidence that PBHs in the mass range $5 \times 10^{17} < M_{\rm
  BH}/{\rm g} < 10^{20}$  cannot constitute a major fraction of dark
matter. This interpretation, however, has been
disputed~\cite{Carr:2020gox}. {\it (iii)}~For $\mbh \sim 5
\times 10^{11}~{\rm g}$, the Schwarzschild radius is $r_s \sim 5
\times 10^{-5}~\mu{\rm m}$, whereas for $\mbh \sim 10^{17}~{\rm g}$, we
have $r_s \sim 2
\times 10^{-2}~\mu{\rm m}$, justifying our assumption that these black
holes are 5-dimensional objects. It is noteworthy that a black hole with $\mbh \sim 1
\times 10^{21}~{\rm g}$ has a horizon radius $r_s \sim 2~\mu{\rm m}$, 
saturating the range of validity of our 5D description. In Fig.~\ref{fig:2} we illustrate how the longer lifetime of PBH
perceiving the mesoscopic-scale extra-dimension modifies the constraints on
$f_{\rm PBH}$. Moreover, the Hawking temperature of the lightest PBHs evaporating today, with $M_{\rm BH} \sim
10^{12}~{\rm g}$, is roughly 1~MeV. For this mass scale, Hawking radiation could
potentially explain the well-known Galactic 511~keV gamma-ray line. It
is this that we now turn to study.

\section{A new PBH window for  explaining the Galactic 511 \lowercase{ke}V line}
\label{sec:4}

It has long been known that electron–positron annihilation proceeds at a surprisingly high rate in the central region of the Galaxy~\cite{Johnson:1972}. In particular, the SPI spectrometer on the INTEGRAL satellite has detected an intense 511~keV gamma ray line flux aligned with the Galactic center (bulge component)~\cite{Knodlseder:2005yq,Jean:2005af}, and also provided evidence for the line in the disk or halo component~\cite{Weidenspointner:2007rs}.

A variety of potential astrophysical sources explaining this signal
have been proposed~\cite{Prantzos:2010wi}, including
PHBs~\cite{Frampton:2005fk,Bambi:2008kx}. The source of positrons
responsible for the 511-keV line must generate ${\cal O} (10^{50})$
positrons per year~\cite{Fuller:2018ttb}. However, if these positrons
are injected at even mildly relativistic energies, higher-energy gamma
rays will also be produced. Diffuse Galactic gamma-ray data strongly
constrain the positron injection energy to be $\alt 3~{\rm
  MeV}$~\cite{Beacom:2005qv}. Another key constraint comes from the local $e^+e^-$ flux as measured by Voyager 1~\cite{Boudaud:2018hqb}.

In the 4D theory the black temperature scales as
\begin{equation}
T_{{\rm BH},4D} \simeq 1.05 \left(\frac{M_{\rm BH}}{10^{16}~{\rm g}} \right)^{-1}~{\rm MeV} \, .
\end{equation}
Herein we are particularly interested in black holes with masses above the evaporation limit. In this mass range,  PBHs are Hawking evaporating today, emitting particles with a characteristic spectrum centered around tens of MeV. Note that most of these PBHs
 would be excluded by Galactic gamma-ray observations.

In the 5D theory discussed herein, PBHs evaporating today are bigger,
longer-lived, and colder than in the 4D theory;
the PBH temperature scales as
\begin{equation}
T_{{\rm BH}, 5D} \sim  \left(\frac{M_{\rm BH}}{10^{12}~{\rm g}} \right)^{-1/2}~{\rm MeV} \, .
\end{equation}

In the most recent data analysis it was shown that in 4D theory
primordial black holes in
mass range of $1 < M_{\rm BH}/10^{16}~{\rm g} < 4$ could potentially
produce the 511~keV gamma-ray signal if  $10^{-4} < f_{\rm PBH}  < 4 \times 10^{-3}$~\cite{Keith:2021guq}. This study takes into account the
gamma-ray fluxes
measured by INTEGRAL in the 0.1-0.2~MeV, 0.2-0.6~MeV, 0.6-1.8~MeV
bands, and by COMPTEL in the 1-3~MeV, 3-10~MeV, 10-30~MeV bands,
as well as the local Voyager constraint on the flux of
$e^+e^-$. Remarkably, the regions of parameter space in which PBHs
could accommodate the observed 511-keV excess require a PBH number density  in the vicinity of the Solar System of
\begin{equation}
n^{\rm local}_{\rm PBH} = \frac{f_{\rm PBH} \ \rho^{\rm local}_{\rm DM}}{M_{\rm BH}} 
\simeq 1.2 \times 10^{-4} \, {\rm AU}^{-3} \times \bigg(\frac{f_{\rm
    PBH}}{10^{-3}} \bigg) \, \bigg(\frac{M_{\rm BH}}{2\times 10^{16} \, {\rm
    g}}\bigg)^{-1} \,,
\label{Dan}
\end{equation}
where $\rho^{\rm local}_{\rm DM} = 0.4 \, {\rm GeV}/{\rm cm}^3$ is the
local dark matter density~\cite{Keith:2021guq}. Since we have seen that black holes in the
5D theory with temperature of 1 MeV have masses roughly 4 orders of
magnitude smaller, to a first approximation a simple rescaling of the
result of  (\ref{Dan}) while demanding the same number density
suggests that an interpretation of the 511-keV line would be, in principle, possible for $f_{\rm PBH} \sim 10^{-7}$. 

As already noted
in~\cite{Keith:2021guq}, for such particular number
density, the closest PBH would be located at a distance
\begin{equation}
d \sim \left(\frac{3}{4\pi n^{\rm local}_{\rm PBH}}\right)^{1/3} \sim \mathcal{O}(10 \,
  {\rm AU}) \, .
\end{equation}
This suggests that the Solar System could contain several hundred black
holes at any given moment and detectability of their Hawking
evaporation could be at reach of future gamma-ray
telescopes~\cite{Keith:2022sow}. Moreover, e-ASTROGAM~\cite{e-ASTROGAM:2016bph}
would not only be able to detect the Hawking radiation from a PBH
population responsible for the 511~keV excess, it would be able to
characterize the properties of such a population with remarkable
precision~\cite{Keith:2022sow}. A simultaneous study of the allowed $(M_{\rm BH}
, f_{\rm PBH})$ parameter space could then be used to
disentangle a PBH evaporating in 4D from one evaporating in 5D. 

\section{Black Hole Growth by Accretion}
\label{sec:5}

In Sec.~\ref{sec:3} we estimated the black hole lifetime assuming that black holes do not accrete. In
general, the net change of the black hole mass is given by
\begin{equation}
\frac{dM_{\rm BH}}{dt} = \left. \frac{dM_{\rm BH}}{dt}\right|_{\rm
  accr} + \left. \frac{dM_{\rm BH}}{dt}\right|_{\rm evap} \,\, ,
\end{equation}
where $dM_{\rm BH}/dt|_{\rm evap}$ is given by Eq. (\ref{dife}) and
\begin{equation}
\left. \frac{dM_{\rm BH}}{dt}\right|_{\rm accr} \approx \pi \left(\frac{n+3}{2}
\right)^{2/(n+1)}  \ \frac{n+3}{n+1} \ r_s^2 \ \varepsilon \,,  
\end{equation}
where $\varepsilon$ is the energy density of the plasma in the
vicinity of the event horizon~\cite{Chamblin:2003wg}. Note, however,
that any correction from the $dM_{\rm BH}/dt|_{\rm accr}$ term will tend to enlarge the
black hole lifetime, and so the conclusions presented in Sec.~\ref{sec:3} would still hold.

The mesoscopic-size dimension imposes generic constraints on the production of PBHs. Namely, as in the context of large extra dimensions~\cite{Arkani-Hamed:1998cxo,Dienes:1998hx}, the universe should remain 4D at the nucleosynthesis MeV temperature, even if the compactification scale is much smaller (at meV). This is attributed to the stabilization of the extra dimension which should happen actually even before the reheating temperature. This of course assumes that the inflation mechanism can be re-adapted in a higher dimensional theory and implements also the stabilization. Hence, the 5D PBHs should be produced after inflation but before reheating.

Since black holes and dark matter are diluted by cosmic expansion in
the same way, in the absence of accretion and decay the
PBH mass function given in (\ref{psiM}) is a constant over time. Although a precise characterization of
$\psi (M_{\rm BH})$ is beyond the
scope of this paper, we note that the bigger, longer-lived, and colder
5D black holes are more prompted to growth through
accretion than those in the 4D theory. As noted
in~\cite{DeLuca:2021pls} PBHs may accrete and increase their mass by
several orders of magnitude. Moreover, accretion may play a critical
role in explaining the LIGO-VIRGO events in terms of PBHs~\cite{DeLuca:2020qqa}.
However, we stress once more that the 5D description breaks down for
$M_{\rm BH} \agt 10^{-12} M_\odot$, and so while the 5D enhanced
accretion effects could influence $\psi(M_{\rm BH})$ in the golden window of
black-hole mass range ($10^{14} \alt M_{\rm BH}/{\rm g} \alt 10^{21}$) where PBHs can account for all of the dark matter content
of the Universe, it will play a negligible role in the mass growth of
$M_\odot$-scale black holes.

\section{Conclusions}
\label{sec:6}

We have studied some phenomenological aspects of black holes
perceiving the dark dimension and analyzed the impact of these higher
dimensional objects in assessing the fraction of dark matter that could
be composed of PBHs. We have shown that 
the rate of Hawking
radiation slows down for 5-dimensional black holes and thereby an
all-dark-matter interpretation in terms of PBHs for 
$10^{14} \alt \mbh/{\rm g} \alt 10^{21}$ should be possible. We have
also shown that an explanation of the Galactic 511-keV
line could be possible for
$M_{\rm BH} \sim 10^{12}~{\rm g}$ if $f_{\rm PBH} \sim 10^{-7}$. Of
course, for a PBH distribution that peaks at $M_{\rm BH} \sim
10^{15}~{\rm g}$, one can, in
principle, obtain a simultaneous all-dark-matter interpretation, with
an explanation of the
Galactic 511~keV gamma-ray signal. These
results are strongly dependent on the choice of $M_{{\rm Pl},n}$ and to
a lesser degree on the behavior of $\sigma_s$ with $Q$. 

It is interesting to note that a rotating black hole would first shed its spin radiating
  particles, dominantly in the equatorial plane. Roughly 25\% of its mass is lost in the so-called ``spin down
phase''~\cite{Giddings:2002kt}. Spin down leaves a
Schwarzschild black hole which continues to Hawking radiate in the
``Schwarzschild phase.'' Since the radiation temperature in the Schwarzschild phase
is larger than the one in
the spin-down phase~\cite{Page:1976ki}, we conclude that existing
limits on spinning PBHs~\cite{Laha:2020vhg} would also be relaxed. All
in all, our 
results are also valid for PBHs produced with angular momentum.

It is
also interesting to note that the black hole decay rate could be slowed down
due to quantum effects, when compared to the semiclassical Hawking
radiation adopted in our calculations~\cite{Dvali:2020wft}. These quantum corrections would become
important if the black hole half-time is comparable to the age of the Universe.

In summary, within the 5D model proposed in~\cite{Montero:2022prj}, with a species scale at
$10^{10}~{\rm GeV}$, PBHs sensing the extra dimension would have a
larger horizon radius, which scales as $M_{{\rm Pl},n}^{-(2+n)/(1+n)}$
as shown in (\ref{Sch}). From (\ref{TBH}) we see that a larger horizon radius in turn implies a smaller
black hole temperature $T_{\rm BH}$. Now, one can check in  (\ref{dife}) that the Hawking radiation
$\propto T_{\rm BH}^2$, and therefore the PBHs in the 5D theory live longer, 
automatically relaxing existing bounds on $f_{\rm PBH}$.  In conclusion, the PBHs as dark matter candidates could provide a  complete picture for the swampland and its cosmology.

\section*{Acknowledgments}

We thank Gia Dvali and Ranjan Laha for discussion, and Sergio Palomares-Ruiz for permission to reproduce Fig.~\ref{fig:1}. The work of L.A.A. is supported by the U.S. National Science
Foundation (NSF Grant PHY-2112527). The work of D.L. is supported by the Origins
Excellence Cluster and by the German-Israel-Project (DIP) on Holography and the Swampland.

\end{document}